# Generalized Fibonacci zone plates

Jie Ke[1,2], Junyong Zhang*, Jianqiang Zhu

[1]Shanghai Institute of Optics and Fine Mechanics, Chinese Academy of Sciences, Shanghai 201800, China
[2]University of Chinese Academy of Sciences, Beijing 10049, China
*Corresponding author: zhangjin829@163.com

We propose a family of zone plates which are produced by the generalized Fibonacci sequences and their axial focusing properties are analyzed in detail. Compared with traditional Fresnel zone plates, the generalized Fibonacci zone plates present two axial foci with equal intensity. Besides, we propose an approach to adjust the axial locations of the two foci by means of different optical path difference, and further give the deterministic ratio of the two focal distances which attributes to their own generalized Fibonacci sequences. The generalized Fibonacci zone plates may allow for new applications in micro and nanophotonics.

doi:10.3788/COLXXXXXX.XXXXXX.

Recently the Fibonacci sequence proposed by Italian mathematician Leonardo de Pisa has been successfully employed in the development of different photonic devices. Photonics is a potential field of applications for novel devices designed and constructed by using a Fibonacci sequence as a consequence of its unique properties. The focusing and imaging properties of Fibonacci optical elements, *e.g.*, quasicrystals [1, 2], gratings [3-5], lenses [6-8], zone plates [9], *etc.*, are studied in detail.

Focusing of soft x-ray [10] and extreme ultraviolet (EUV) has many applications in physical and life sciences, such as high-resolution microscopy, spectroscopy, and lithography [11]. Traditional Fresnel zone plates (TFZPs), which have inherent limitations [12, 13], can be used for this kind of focusing [14, 15]. Some aperiodic zone plates, generated with the fractal Cantor set, have been proposed to overcome some of these limitations [16, 17], another interesting mathematical generator of aperiodic zone plates is the aforementioned Fibonacci sequence. In mathematics, many mathematicians have extensively studied the Fibonacci sequence and its various generalizations [18-22] in the past decades.

In this letter, we introduce the aperiodic generalized Fibonacci sequences into the zone plates, and come to a conclusion that the generalized Fibonacci zone plates (GFiZPs) can yield two equal axial intensity foci with adjustable location.

Let us review the standard Fibonacci sequence, which is defined by the follow recurrence relation

$$F_1 = F_2 = 1, \; F_{n+2} = F_{n+1} + F_n \; (n \in N^+). \quad (1)$$

Obviously, $x_1 = (1+\sqrt{5})/2$ and $x_2 = (1-\sqrt{5})/2$, which are associated with the classical geometrical problem of the golden section, are the characteristic roots of the characteristic equation $x^2 - x - 1 = 0$. The golden mean is defined as the limit of the ratio of two consecutive Fibonacci numbers:

$$\psi = \lim_{j \to \infty} F_j / F_{j-1} = (1+\sqrt{5})/2. \quad (2)$$

Similarly, we can extend the standard Fibonacci sequence to the generalized Fibonacci sequences via the following initial seed elements

$$F_j = a_j \; (j, a_j \in N^+). \quad (3)$$

And the corresponding linear recursion relation of the generalized Fibonacci sequences can be written as

$$F_n = \sum_{m=1}^{j} C_m F_{n-m} \; (C_m \in R, n, j \in N, n > j \geq 2). \quad (4)$$

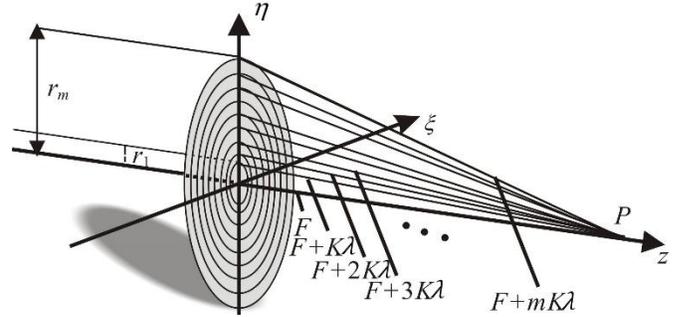

Fig. 1. Different optical path between zones under a plane wave incidence.

We now retrospect the design of TFZPs based on the plane wave incidence as shown in Figure 1. All the designed rays are converged upon a single point, the expected focus P. As known, the half wave zones of TFZPs can be determined by [14]

$$\sqrt{r_m^2 + F} - \sqrt{r_{m-1}^2 + F} = \lambda/2, \quad (5)$$

where $r_m$ denotes the radius of the mth zone, $F$ is the expected focal distance and $\lambda$ is the incident wavelength. Suppose we alter the optical path difference between two adjacent zones as $K\lambda$ ($K \in R^+$) instead of $\lambda/2$, then one can further describe the radii of zones as

$$r_m = \sqrt{(mK\lambda)^2 + 2mK\lambda F}, \qquad (6)$$

where $K$ is the optical path difference parameter (OPDP) in this letter, whose influence on axial bifocal locations of GFiZPs will be investigated later.

GFiZPs can be generated similar to the process of TFZP. Taking a generalized Fibonacci sequence into account, whose initial seed elements are $F_1 = 1$, $F_2 = 2$ and $F_3 = 3$ and the recursion relation is $F_{n+3} = F_{n+2} + F_{n+1} + F_n$, after encoding three seed elements as $(F^1, F^2, F^3) = (0, 01, 011)$, the six-order switching sequence $F^6$ is 01101001101011010011 while 1 denotes transparent zones and 0 denotes opaque ones. That means the number of total zones is 20 and the number of transparent zones is 11 shown in Figure 2.

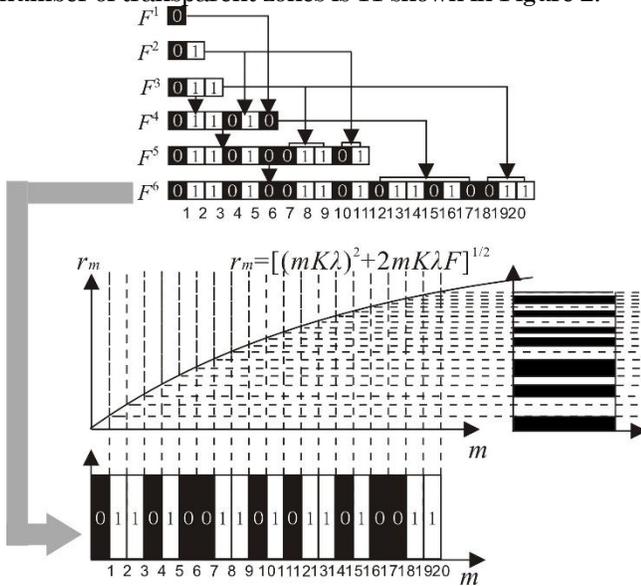

Fig. 2. Flow chart of generation of GFiZPs.

Theoretically, the diffraction field of the GFiZPs and the associated Fresnel zone plates can be numerically calculated by the Rayleigh-Sommerfeld diffraction integral formula under the condition of a plane wave incidence with unit amplitude [23-25]

$$U(x, y, z) = \iint_\Sigma t(\xi, \eta, 0) \frac{\exp(ikR)}{i\lambda R}(1 + \frac{i}{kR})\frac{z}{R}\mathrm{d}\xi \mathrm{d}\eta, \qquad (7)$$

where $t(\xi, \eta, 0)$ is the pupil function of GFiZPs, $i$ is the imaginary unit, $k$ is the wave number, $z$ is the axial distance from the pupil plane, and $R$ denotes the distance between point $(\xi, \eta, 0)$ and point $(x, y, z)$.

To investigate the axial focusing performance of GFiZPs, a typical kind of GFiZPs of ten-order switching sequence $F^{10}$ is shown in Figure 3(a), and the corresponding simulation parameters are as follows: $\lambda = 632.8nm$, $(F^1, F^2, F^3) = (0, 01, 011)$, $F = 4cm$, $K = 0.5$ and the corresponding linear recursion relation is $F_{n+3} = F_{n+2} + F_{n+1} + F_n$, whose characteristic equation is $x^3 - x^2 - x - 1 = 0$, and the characteristic roots are 1.839, $-0.420 + 0.606i$, and $-0.420 - 0.606i$, respectively. For comparison, the TFZPs with the same resolution are represented in Figure 3(b). The number of transparent zones is 125 for GFiZPs and 115 for TFZPs while each of them has a total of 230 zones. The axial normalized intensity computed for GFiZPs and the associated TFZPs are shown in Figure 3(c). Obviously, in this case, the first focus of the GFiZPs is located at $f_1 = 3.086cm$ and the other one at $f_2 = 5.682cm$ while the prime focal distance of TFZPs is $4.000cm$. Thus, the ratio of the two focal distances satisfies $f_2/f_1 = 1.841$. However, because of the diversity of the GFiZPs based on different encoded seed elements, they may present more than two foci under a plane wave illumination. But in this letter, we just consider the bifocal focusing properties.

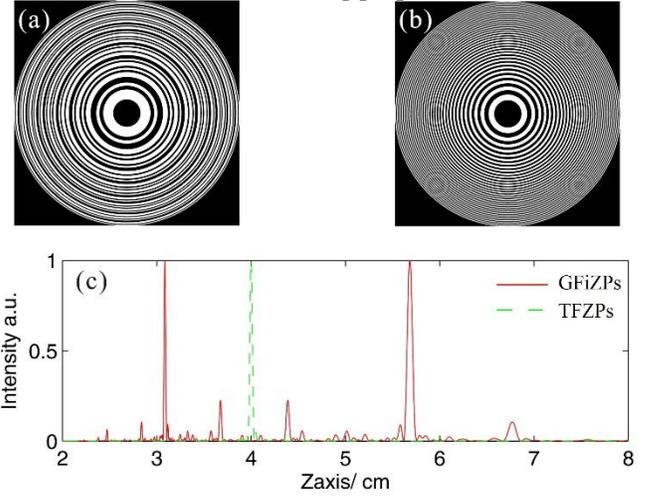

Fig. 3. (a) GFiZPs ($F^{10}$ in this case); (b) TFZPs with the same resolution; (c) Normalized intensity distribution along the optical axis produced by GFiZPs and TFZPs.

Apart from the value $K$, the other parameters remain the same value above. In this case $K$ is equal to 0.45, the two focal distances are changed to $2.777cm$ and $5.114cm$ shown in Figure 4(a). Other results are shown in Figure 4 (b) ~ 4(e).

Table 1. Locations of on-axis foci with different OPDP

| OPDP ($K$) / Focal distances | 0.45 | 0.50 | 0.85 | 1.54 | 2.38 |
|---|---|---|---|---|---|
| $f_1 (cm)$ | 2.777 | 3.086 | 5.252 | 9.53 | 14.76 |
| $f_2 (cm)$ | 5.114 | 5.682 | 9.667 | 17.54 | 27.15 |

From Table 1 we know that the ratio of the two focal distances remains the same while their absolute locations are adjustable due to the different OPDP. What's important is that amplification or constriction factor of two focal distances is exactly equal to the ratio of the two values of OPDP, such as $2.777/5.252 \approx 0.45/0.85$.

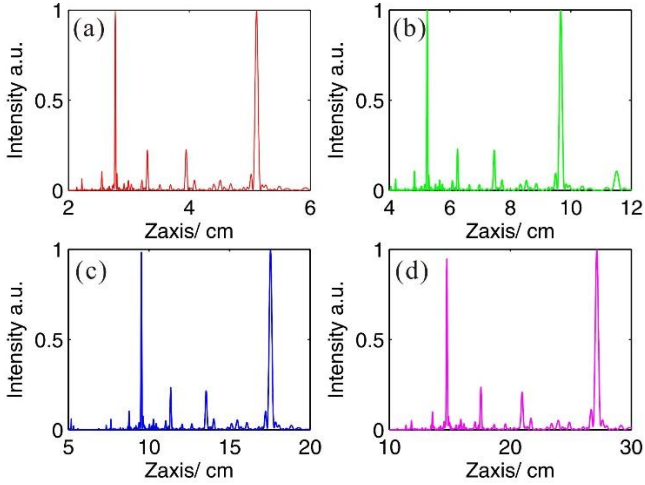

Fig. 4. Normalized intensity distribution along the optical axis produced by GFiPS ($F^{10}$ in this case) with different OPDP: (a) $K = 0.45$; (b) $K = 0.85$; (c) $K = 1.54$; (d) $K = 2.38$.

Besides, the two focal distances as a function OPDP are plotted in Figure 5. The two focal distances are directly proportional to the OPDP. The ratio of two OPDP is equal to that of two focal distances. Hence, in this way, one can regulate and control the two focal distances to a certain degree.

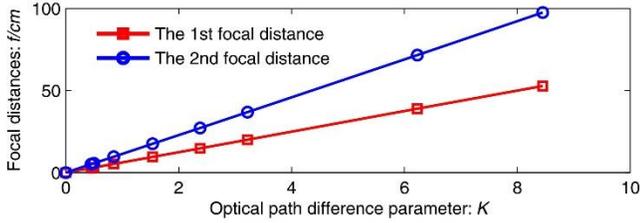

Fig. 5. The relation of focal distances and values of OPDP.

The Fibonacci zone plates can present two equal intensity foci and the ratio of the two focal distances approaches the golden mean [7] which is a fixed value. But for GFiZPs, the ratio may be adjustable due to the different generalized Fibonacci sequences.

For the purpose of adjustable ratio of the two axial focal distances, other four kinds of GFiZPs are investigated. The parameters are as follows: $\lambda = 632.8nm$, $F = 4cm$, $K = 0.5$. For two seed elements, they are $F_1 = 2$ and $F_2 = 3$ and coded as $(F^1, F^2) = (01, 010)$; while for three seed elements, they are $F_1=1$, $F_1=2$ and $F_2=3$ and coded as $(F^1, F^2, F^3) = (0, 01, 101)$. Figure 6 shows axial intensity distribution produced by GFiZPs based on different linear recursion relations, which are $F_{n+3} = -F_{n+2}+F_{n+1}-F_n$, $F_{n+2} = -F_{n+1}+F_n$, $F_{n+2} = 2F_{n+1}+0.25F_n$ and $F_{n+2} = -2F_{n+1}+0.25F_n$, respectively. Here negative sign means complement operation. The ratio of the two focal distances changes from 1.841 to 1.618 and 2.116 due to the different generalized Fibonacci sequences.

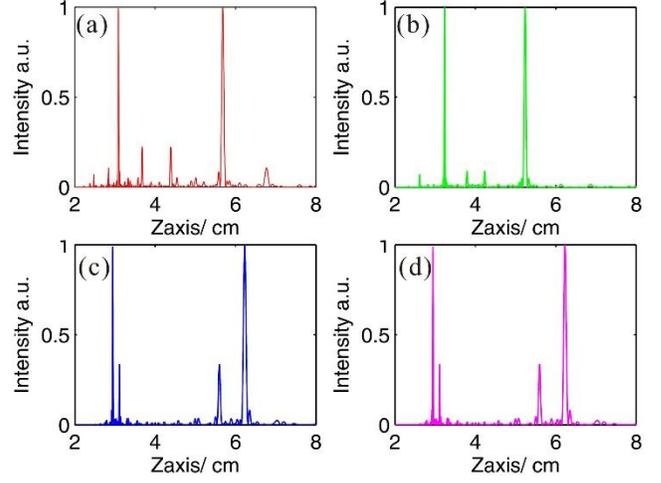

Fig. 6. Normalized axial intensity distribution produced by GFiZPs based on different generalized Fibonacci sequences.: (a) $F_{n+3} = -F_{n+2}+F_{n+1}-F_n$ ($F^{10}$ in this case); (b) $F_{n+2} = -F_{n+1}+F_n$ ($F^{11}$ in this case); (c) $F_{n+2} = 2F_{n+1}+0.25F_n$ ($F^8$ in this case); (d) $F_{n+2} = -2F_{n+1}+0.25F_n$ ($F^8$ in this case).

Table 2. Locations of on-axis foci with different GFiZPs

| Sequences (Total zones) | $f_1(cm)$ | $f_2(cm)$ | $f_2/f_1$ |
|---|---|---|---|
| $F_{n+3} = F_{n+2}+F_{n+1}+F_n$ (230) | 3.086 | 5.682 | 1.841 |
| $F_{n+3} = -F_{n+2}+F_{n+1}-F_n$ (230) | 3.086 | 5.682 | 1.841 |
| $F_{n+2} = -F_{n+1}+F_n$ (233) | 3.235 | 5.238 | 1.618 |
| $F_{n+2} = 2F_{n+1}+0.25F_n$ (237) | 2.944 | 6.229 | 2.116 |
| $F_{n+2} = -2F_{n+1}+0.25F_n$ (237) | 2.944 | 6.229 | 2.116 |

Table 2 shows the value of focal distances. It is necessary to point out that the ratio of two focal distances of GFiZPs generated by the generalized Fibonacci sequences with the proper encoding can be equal to one of the characteristic roots. More important, it suggests that the ratio of two focal distances is also adjustable and just relevant to the given switching sequence.

In conclusion, we proposed a new family of zone plates whose structure is based on the generalized Fibonacci sequences, and investigated the axial focusing properties of the GFiZPs. Due to the complexity and diversiform design for GFiZPs, it may present more than two foci under a plane wave incidence, but we just studied the bifocal focusing properties in this letter. It suggested that GFiZPs presented two axial foci with equal intensity. Not only the absolute locations of the two foci are adjustable by use of different optical path difference parameter, but also the ratio of two focal distances is altered as required by use of different generalized Fibonacci sequences. Apart from the applications in the field where TFZPs are usually applied, such as terahertz (THz) imaging [26] and x-ray microscopy [27], the proposed method of designing the structure of GFiZPs can offer reference for photon sieves [28-30].

This work was supported by the National Natural Science Foundation of China (No. 61205212 and 61205210).